\documentclass{jaa}
\usepackage{natbib}
\usepackage{graphicx}
\usepackage{color}

\usepackage{hyperref}
\hypersetup{colorlinks,linkcolor={blue},citecolor={blue},urlcolor={red}} 

\def\aap{A\&A}%
\def\aapr{A\&A~Rev.}%
\def\aaps{A\&AS}%
\def\aj{AJ}%
\def\araa{ARA\&A}%
\def\apj{ApJ}%
\def\apjl{ApJ}%
\def\apjs{ApJS}%
\def\mnras{MNRAS}%
\def\nat{Nature}%
\def\pasa{PASA}%
\def\pasj{PASJ}%
\def\pasp{PASP}%
\def\ssr{Space~Sci.~Rev.}%
\begin{document}\sloppy

\title{The BL Lac Object OJ 287: Exploring a Complete Spectrum of Issues\\ 
Concerning Relativistic Jets and Accretion}


\author{Pankaj Kushwaha\textsuperscript{1,2,$\dagger$},*}
\affilOne{\textsuperscript{1}Aryabhatta Research Institute of Observational Sciences
(ARIES), Nainital 263001, India.\\}
\affilTwo{\textsuperscript{2}Department of Physical Sciences, Indian Institute of
Science Education and Research Mohali, Knowledge City, Sector 81, SAS Nagar, Punjab
140306, India.\\}
\affilThree{\textsuperscript{$\dagger$}Aryabhatta Postdoctoral Fellow.}


\twocolumn[{

\maketitle

\corres{pankaj.kushwaha@iisermohali.ac.in}


\begin{abstract}
The BL Lacertae  (BL Lac) object OJ 287 is one of the most dynamic blazars across
the directly accessible observational windows: spectral, timing, polarization,
and imaging.  Apart from behaviors considered characteristics of blazars, it 
exhibits peculiar timing features like quasi-periodicity in optical flux  as well
as radio-detected knots position and has shown diverse transient spectral features
like  a new broadband continuum dominated activity phase, Seyfert-like soft-X-ray
excess,  highly transient iron line absorption feature, a thermal-like 
continuum-dominated optical phase, large  optical polarization swings associated with one of the timing features, etc. that are rare in blazars and contrary to
currently prevailing view of BL Lacs. Theoretical considerations, supported by
existing observations invoke scenarios involving a dynamical interplay of accretion
and/or strong-gravity-induced events (tidal forces) in a binary supermassive black
hole (SMBH) scenario to impact-induced jet and only jet activities.  Many of these scenarios have some definite and quite distinctive observationally testable  predictions/claims. 
These considerations make OJ 287 the only  BL Lac to have an activity
phase with dominance related to accretion and/or accretion-perturbation-induced jet 
activities. We present a brief overview of the unique spectral features and discuss
the  potential of these features in exploring not only relativistic jet physics but
issues  pertaining to accretion and accretion-regulated jet activities,
i.e. the whole spectrum of issues related to the jet-accretion paradigm.
\end{abstract}


}]


\doinum{12.3456/s78910-011-012-3}
\artcitid{\#\#\#\#}
\volnum{000}
\year{0000}
\pgrange{1--}
\setcounter{page}{1}
\lp{1}

\section{Introduction}
Astrophysical jets are a highly collimated outflow of material/plasma and large-scale
jets, extending far beyond the sphere of influence of the source, though scarce,
are a relatively common phenomenon associated with accreting systems. They have
been observed in astrophysical systems of all masses, from stars \citep[e.g.][]
{2018A&ARv..26....3A}, X-ray binaries \citep[e.g.][]{2016ApJS..222...15T}, to
galaxies \citep[e.g.][]{2019ARA&A..57..467B}. For a successful outflow, the ejected
material must leave the gravitational influence of the source, and thus, jets
associated with highly compact objects are relativistic, and observations
report/indicate a diverse range of jets -- mild to relativistic, (mostly) transient
jets in micro-quasars \citep[e.g.][]{2016ApJS..222...15T}, persistent relativistic
jets in active galaxies \citep[AGNs; e.g.][]{2019ARA&A..57..467B}, and highly
transient but ultra-relativistic in gamma-ray bursts \citep[GRBs; e.g.][]{2014PASA...31....8G}.
The emission is also extremely diverse, with both thermal and non-thermal components,
often showing dynamic co-evolution at different activity phases. Although almost
every aspect related to relativistic jets and accretion, from jet triggering mechanisms
to acceleration of particles to relativistic energies and broadband emission 
is yet to be understood \citep[e.g.][]{2017SSRv..207....5R}, the apparent exclusive
association of large scale jets with accreting systems, nonetheless, 
strongly indicates accretion as a primary driver while the observed diversity
implies a highly dynamic and complex multi-scale physics. The other ubiquitous ingredient
indicated by the observations is magnetic fields \citep[e.g.][]{2014Natur.510..126Z}.

The central region incorporating accretion and jet launching is extremely compact
such that even for the nearest sources with the best existing resolution, e.g. the
{\it Event Horizon Telescope} \citep[EHT;][]{2019ApJ...875L...1E}, it is still beyond
our resolution scale. This leaves only spectral, timing, polarization, and imaging
(to a limited extent; e.g. superluminal and large-scale features) as the direct ways
to peek into these sources and derive/infer scales as well as dynamics. Thus, in this
overall accretion-jet paradigm, sources where accretion-powered emission
primarily dominates the observed features represent the one end and the
ones where jet emission dominates represent the other end,  while  sources
where both can be seen during some activity phases and evolve to either end of the spectrum are ideal for investigating and understanding the missing links.

The most powerful accretion-powered sources in the universe are active
galactic nuclei  (AGNs) and a small fraction of these, designated radio-loud, hosts the
most powerful, persistent large-scale relativistic jets. Blazar is the AGNs subclass
in which a jet points nearly at the Earth. They emit a featureless continuum that
spans the entire accessible electromagnetic (EM) spectrum from radio to GeV/TeV
gamma-rays \citep[e.g.][]{2015ApJ...807...79H,2018A&A...620A.181A}, characterized
 by a broad double-humped spectral energy distribution (SED) and high and rapid variability  of continuum and polarization on all feasible timescales
being probed so far \citep[e.g.][]{2018ApJ...863..175G}. They are the most powerful
persistent broadband emitter in the universe, and the radiation in every energy band
is dominated almost
entirely by the jet emission except for a few exceptions \citep[e.g. PKS 1222+216;][]{2011A&A...534A..86T,2014MNRAS.442..131K}. Even for 
most of these exceptional sources, the jet emission outshines  other emission
features during the bright activity phases \citep[e.g. 3C 454.3;][]{2007A&A...473..819R}. 

The dominance of jet emission in all the accessible EM bands makes blazars the ideal
sources to explore, understand, and characterize relativistic jet emission, thereby
offering a direct peek into jet physical conditions and extreme physics. However,
the outshining of other likely emission components -- either galaxy or various 
ingredients of accretion also presents formidable challenges vis-a-vis comparative
studies with other accretion-powered sources where both thermal – widely accepted
to be associated with accretion and non-thermal broadband emission – taken as the
signature of jet emission is seen. This limits us to far fewer options and primarily
to statistical inferences from studies of a sample of sources for comparative  views.
However, more than often, the outcomes from different approaches are conflicting. For
example, from a spectral point of view, blazars are considered extreme AGNs with
the jet continuum dominating in entirety, while from a flux variability point of
view over the long-term, they appear similar to other accretion-powered sources
\citep[e.g.][]{2016ApJ...822L..13K,2017ApJ...849..138K}. At short timescales
(minutes to hours), the trends are unclear \citep[e.g.][]{2020Galax...8...66K,
2020ApJ...897...25B}. Similarly, for some blazars, spectral classification based
on SED is very different  from that of the kinematic studies at radio, e.g. OJ 287
\citep{2016A&A...592A..22H}.

The observed behavior of blazars covers an enormous range in each of direct observational
windows -- emission spread over the entire EM spectrum from radio to GeV/TeV gamma-rays
($\rm\gtrsim 17-20$ orders of magnitude), flux variability on minutes and even less
to decades and more \citep[$\rm \gtrsim 6-7$ orders of magnitude in timing; e.g.][]
{2018ApJ...863..175G}, polarization degree (PD) variation from 0 to 50\% while
polarization angle (PA) rotation  of $360^\circ$ and more \citep[e.g.][]{2016A&A...590A..10K,
2010ApJ...710L.126M}, and jet extending from the unresolvable compact site, probably
even compact than our Solar system up to galaxy cluster scales (Mpc)
covering $\rm \gtrsim  10-24$ orders of magnitude in spatial scale \citep[e.g.][]
{2006ApJ...648..910U,2011ApJ...729...26M}. The enormous range and the requirement of
contemporaneous coverage across the EM bands not only posses apparently insurmountable
challenges for observations but  also theoretical investigations, requiring
prohibitively huge computational resources. Even for the best cases where, to a zeroth
level, the problem can be reduced to one parameter for some specific issues,
coupling it with radiation immediately makes the problem wide open with far too many parameters \citep[e.g.][]{2019ApJ...875L...5E}.

Notwithstanding the resource issue, the continued push and dedicated efforts
with studies exploiting simultaneous\footnote{Note the ``simultaneous'' here is a
misnomer but often used in the literature. Different sensitivities in different
energy bands bars a truly simultaneous multi-wavelength observation.}/contemporaneous
multi-wavelength (MW) data, mostly a Fermi observatory initiated effort\footnote{\url{https://fermi.gsfc.nasa.gov/ssc/observations/multi/programs.html}}, has revealed multitudes
of trends from widely different observations, exploiting widely different methodologies
that are now considered characteristic features of blazars and have been exploited
extensively to explore diverse issues concerning the jet physics \citep[e.g.][and
references therein]{2013ApJ...768...54B,2019MNRAS.482L..80B,2013ApJ...763..134F,
2019NatAs...3...88G,2014Natur.515..376G,2017MNRAS.469..255G,2016A&A...592A..22H,
2011ApJ...727...21J,2014ApJ...780...87M,2017Natur.552..374R,2020ApJ...893L..20L,2021icrc.confE..804R,2015MNRAS.447...36P,2012MNRAS.423..756P,2015MNRAS.451..927Z,
2018ApJ...862L..25Z,
2014Natur.510..126Z}. These trends/characteristic features include -- primarily
stochastic flux variability \citep[e.g.][]{2014ApJ...786..143S,2018ApJ...863..175G}
and similarity of statistical properties with accretion-powered sources in general
\citep[e.g.][]{2016ApJ...822L..13K,2017ApJ...849..138K,2018MNRAS.480L.116S,
2020MNRAS.497.1294T}, intra-night variability \citep[e.g.][]{2021ApJ...909...39G,
2018ApJ...854L..26S,2021Univ....7...15L}, a broad double-humped spectral energy
distribution (SED) with a highly stable location of the two peaks despite strong
variations in flux as well as in Doppler boost, superluminal bright features
\citep[e.g.][]{2013AJ....146..120L,2016A&A...592A..22H}, inverted/flat radio spectra
(MHz -- GHz), usually high PD during flares, etc. There are exceptions to all these,
but only a few. Polarimetric studies though are not as exhaustive as temporal and
spectral studies but the RoboPol\footnote{\url{https://robopol.physics.uoc.gr/}}
led systematic studies have made invaluable contributions to the general polarimetric
behaviors; indicating some trends in polarization, especially a slow PA rotation
during the brightest gamma-ray flares \citep[and references therein]{2018MNRAS.474.1296B,
2019Galax...7...46B}.

Given the extreme and apparently insurmountable requirements, variability in the
directly accessible observables (spectral, flux, polarization, and imaging) is the
only way to explore, infer, and understand these sources. Though blazars are now
firmly well known for flux and polarization variability, concurrent strong spectral
changes indicating new emission components or drastic change in the spectral state are
extremely rare, seen only in a few of the blazars and that too for a relatively very
short duration e.g. Mrk 501 \citep[e.g.][]{1998ApJ...492L..17P,2018A&A...620A.181A},
3C 279 \citep[e.g.][]{2015ApJ...807...79H}, etc. On the contrary, OJ 287 is the
only blazar currently with a  history of strong spectral changes persisting for
comparatively much longer duration \citep[$\rm\gtrsim 4$-year; e.g. ][]{2017IAUS..324..168K,
2017ICRC...35..650B,2018MNRAS.473.1145K,2018MNRAS.479.1672K,2018MNRAS.480..407K,
2020MNRAS.498L..35K,2021ApJ...921...18K,2021MNRAS.508..315P,2021A&A...654A..38P,
2022MNRAS.509.2696S,2021icrc.confE..644K}. It also has been claimed to show a few  recurring
timing features both in flux \citep{1988ApJ...325..628S,2018ApJ...866...11D} as
well as radio images \citep{2017Galax...5...12C,2018MNRAS.478.3199B}  which is contrary
to the general stochastic flux variability. Additionally, a few of the recently
reported/discovered spectral features challenges our widely accepted view of BL Lac
sources, e.g. break in the NIR-optical spectrum from its well-known (smooth) power-law
form indicating a thermal component \citep{2018MNRAS.473.1145K,2020MNRAS.498.5424R},
Seyfert-like soft X-ray excess \citep{2020ApJ...890...47P}, a highly transient iron
line absorption feature indicating relativistic outflow \citep{2020MNRAS.498L..35K}.
The plausible implications of these features encompass every issue from accretion
dynamics to energization of particles to ultra-relativistic energies and broadband
emission, as elaborated and discussed in the following sections.

 Observationally, the biggest advantage in the case of OJ 287, however, is the
 close coincidence of the spectral changes with the $\sim 12$-yr quasi-periodic
optical outbursts (QPOOs) and, hence, the predictability of
expected sighting of these peculiar features, making coordinating MW monitoring
relatively much easier with drastically fewer efforts when viewed in the context
of challenges that plague observations and studies of transients. As stated above
and elaborated below, these make OJ 287 the only blazar with much broader potential
compared to a chosen few  -- suitable for exploration of some specific aspects
of jet physics.  In the next section,  we briefly present the reported peculiar observational features with some comments on the models of QPOOs. We then 
focus on the exhibited spectral changes and argues their potential in exploring
aspects of accretion (based on proposed scenarios in the literature) and jet 
physics -- emission mechanisms, location of emission region, particle spectrum and constraints on highest energies from the optical-UV spectrum, and an outline of how these inputs further allow probe of other issues of the jet-accretion paradigm in section \S\ref{sec:physics} We finally summarised and conclude in section \S\ref{sec:sumCon}
 For an overview of the general observation behavior of OJ 287 across the directly accessible observational windows, we refer to our previous work -- \citet{2020Galax...8...15K}.


\begin{figure*}[!ht]
\includegraphics[scale=1.1, angle=0]{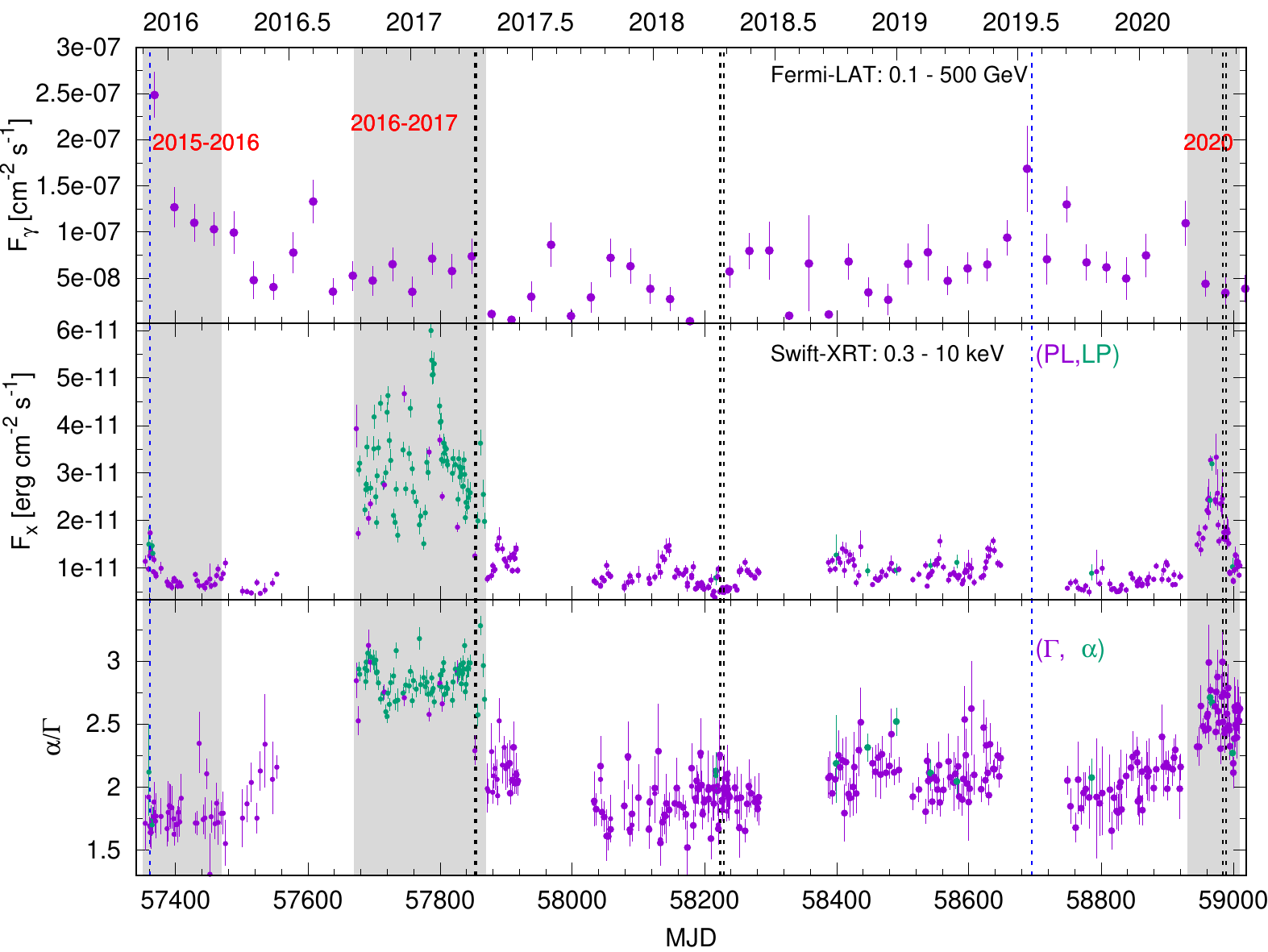}
\caption{Monthly binned gamma-ray light curve from {\it Fermi}-LAT along with X-ray
light curve and the corresponding best-fit spectral indices of OJ 287
from {\it Neils Gherel Swift Observatory}. The shaded regions mark the spectrally
distinct high MW activity periods while the non-shaded parts trace the spectral 
transition and its evolution. The blue vertical dashed lines are the claimed
time of the impact flares \citep{2016ApJ...819L..37V,2020ApJ...894L...1L} while the
black vertical dashed lines represent long-exposure observations by the {\it AstroSat} 
\citep{2022MNRAS.509.2696S,2021icrc.confE..644K}.}
\label{fig:fig1}
\end{figure*}

\section{OJ 287} \label{sec:oj287}
OJ 287 is a BL Lacertae (BL Lac) type object located at a cosmological redshift of z=0.306 \citep{1985PASP...97.1158S,2010A&A...516A..60N}. The BL Lac designation is
the AGN classification scheme based on the strength of the emission lines  with respect
to the  underlying continuum and is attributed to those showing very weak or a complete
absence of emission line features \citep[equivalent width $\rm <~5~{\AA}$;][]{1991ApJ...374..431S} but exhibits unusually high variations in both flux
and polarization and have core-dominated inverted radio spectra. For OJ 287, emission lines have only been seen during its low optical flux states and these few observations suggest strong flux variations \citep{1989A&AS...80..103S,1985PASP...97.1158S,2010A&A...516A..60N,2021ApJ...920...12H}.

OJ 287 is one of the best-explored sources, primarily due to its conducive location
combined with observationally favorable features like high radio and optical brightness,
highly dynamic and correlated MW variability, etc., making it the candidate source
for characterization of the BL Lacertae class of sources \citep{1985PASP...97.1158S}.
This in turn has culminated in one of the richest sets of data among blazars across
the EM spectrum over a diverse range of timescales, taken either individually or in a coordinated fashion. Amongst blazars, it has the longest existing optical data going
back to around 1890 \citep{1973ApJ...179..721V,1988ApJ...325..628S,2013A&A...559A..20H}.

In terms of blazars'  SED-based classification, OJ 287 is categorized as a 
low-frequency/low-energy peaked blazar \citep{2010ApJ...716...30A}. Early MW studies
employing contemporaneous data do not show any appreciable change in the broadband
 SED state during high MW activity phases \citep[e.g.][]{2013MNRAS.433.2380K,
2009PASJ...61.1011S}. However, the MW activities since the end-2015 \citep[e.g.][]{2017MNRAS.465.4423G,2018MNRAS.473.1145K} to date turned out to be very different,
especially in terms of spectral 
changes. An X-ray and MeV-GeV gamma-ray flux evolution of this duration, along with the
best-fit X-ray spectral indices, are shown in Figure \ref{fig:fig1}.  Since X-ray
spectra of OJ 287 have often shown a significant departure from a simple power-law description
\citep[e.g.][]{2018MNRAS.473.1145K,2020ApJ...890...47P}, we used both -- a power-law
and a log-parabola model where the latter can capture this departure. The best-fit 
model was chosen on the basis of the F-test statistics. A strong spectral evolution at X-ray
energies is visible in Figure \ref{fig:fig1} and the shaded regions mark these peculiar activity episodes.

\begin{figure*}[!ht]
\begin{center}
\includegraphics[scale=0.55, angle=0]{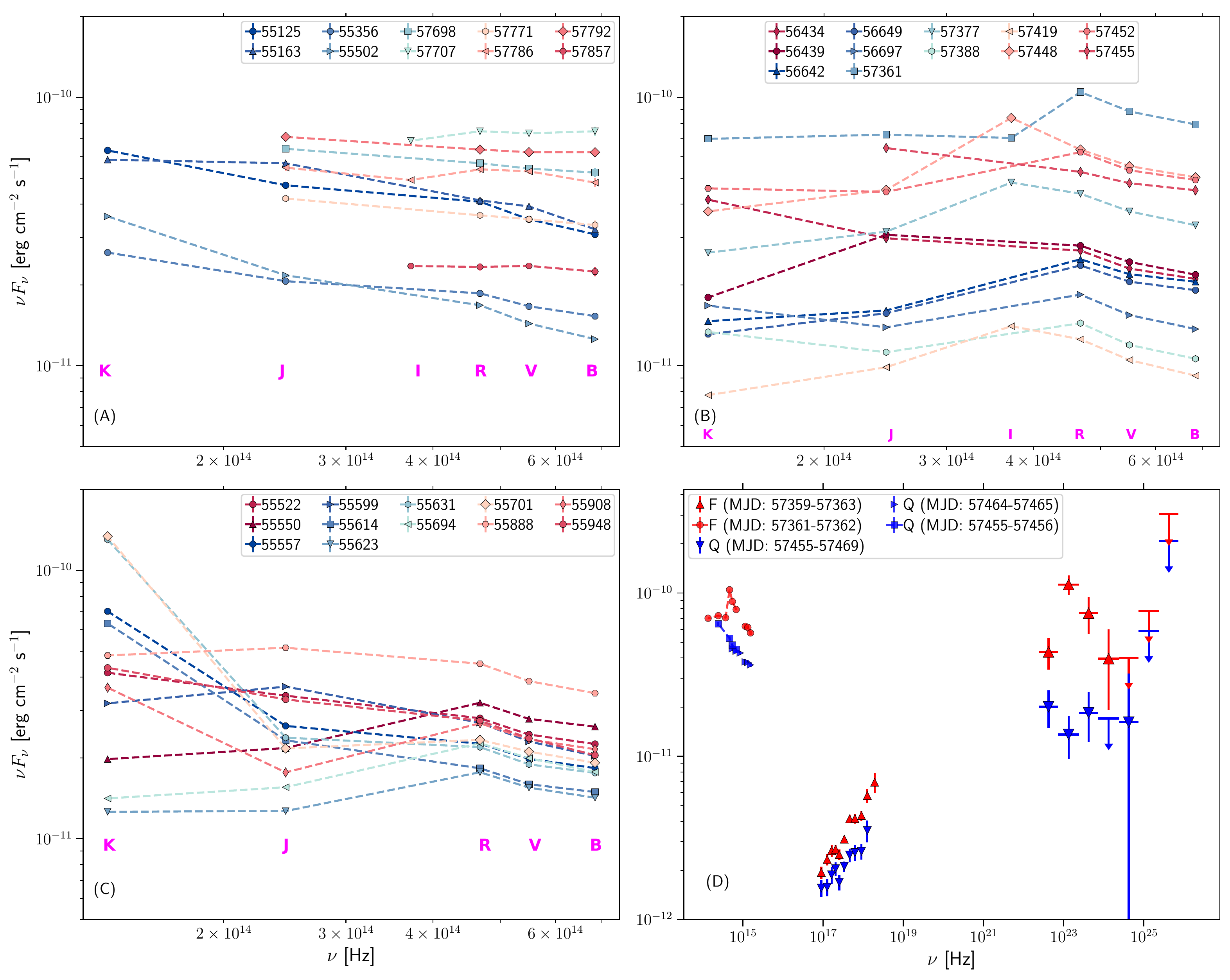}
\end{center}
\caption{A glimpse of the diverse range of NIR-optical spectral states exhibited by 
OJ 287 to date. {\bf (A):} The NIR-optical spectra considered generic of the source -- a
simple power-law spectrum with/without smooth curvature at the either or both the ends. {\bf(B):} The departure of the NIR-optical spectrum from it well-known simple power-law spectrum.  {\bf (C):} An strongly odd variability only in the NIR-K band data from the
SMARTS facility for an extended duration (MJD: 55500 -- 55715), preceding the appearance of the NIR-optical spectral break.  This is likely artificial as NIR data from INAOE, Mexico do not support this \citep{2022ApJS..260...39G}. {\bf (D):} The broadband SEDs from a flaring
(F) and a quiescent (Q) state of the source from its 2015 -- 2016 activity. The
flaring NIR-optical spectrum signifying why we termed the NIR-optical spectrum
departure from a powerlaw form as break -- too sharp compared to the general broadband spectrum of the source/blazars -- closer to thermal emission spectrum \citep{2018MNRAS.473.1145K,2020MNRAS.498.5424R}.  Magenta colored labels (KJIRVB) in panels (A), 
(B), and (C) mark the optical-NIR filters.}
\label{fig:fig2}
\end{figure*}

 MW data of the 2015--2016 activity revealed a sharp spectral break in the NIR-optical
spectrum from its well-known (smooth) power-law  form. A concurrent hardening of 
the MeV-GeV spectrum, as well as a shift in the location of the peak,  was also
observed \citep[ref Figs. \ref{fig:fig2} and \ref{fig:fig3}][]{2018MNRAS.473.1145K}.
 Soon after this, a new MW activity (Figure \ref{fig:fig1}: 2016 -- 2017) with
strong optical to X-ray variations \citep[e.g.][]{2017IAUS..324..168K,2020MNRAS.498L..35K,
2018MNRAS.479.1672K}  was reported, and detailed exploration revealed it to be due to the presence of a  new additional HBL-like component \citep[e.g.][]{2018MNRAS.479.1672K,
2021ApJ...921...18K,2022MNRAS.509.2696S,2021icrc.confE..644K}.  This new state
again reappeared (Figure \ref{fig:fig1}: 2020) in a slightly weaker form in 2020 \citep{2020MNRAS.498L..35K,2021ApJ...921...18K,2022MNRAS.509.2696S}. Furthermore, 
an absorption feature (highly transient) in the X-ray spectrum during the 2020 activity,
\citep{2020MNRAS.498L..35K}, a spectral cut-off in the high-energy end of the optical
synchrotron spectrum during and after 2016--2017 activity \citep{2022MNRAS.509.2696S,
2021icrc.confE..644K}, and a Seyfert-like soft X-ray excess before the 2015--2016
activity \citep{2020ApJ...890...47P} has been reported.

 Altough we have only recently had a quite dense follow-up of the $\sim$12-yr QPOOs
across the EM bands, the close coincidence of the spectral changes reported above
indicates a connection between the two. These provide strong constraints to the
models proposed for the QPOOs when combined with the general behavior of the source
as well as the BL Lacs. In short, based on the 
reported NIR-spectral break and its time of appearance, \citet{2020Galax...8...15K}
argues that \citet[see \citet{2018ApJ...866...11D} for the latest iteration of
the model]{1996ApJ...460..207L} model that invokes the impact of secondary SMBH ($\rm
1.5\times 10^{8}~ M_\odot$) on the accretion disk of the primary ($\rm 1.8\times 10^{10}
~ M_\odot$) for the  QPOOs is broadly favored over the simple
jet precession interpretations.  This and the claims of the high optical to X-ray 
MW activity driven by new HBL-like broadband as a likely tidal-disruption
event \citep[TDE;][]{2021ApJ...920...12H} allows one to explore accretion and to a
limit, connection with the jet in addition to the relativistic jet physics for which
blazars are well known.

Regarding the $\sim12$-yr QPOOs, it should be noted that studies employing diverse
and state-of-the-art timing methodologies \citep{2020ApJ...896..134P,2018ApJ...863..175G}
dispute the  QPOO feature. However, observationally, flares in optical
bands have been observed around the predicted times \citep{2016ApJ...819L..37V,
2020ApJ...894L...1L,2018ApJ...866...11D}. 

 
\section{OJ 287 Spectral Changes: Exploring Jet-Accretion Paradigm in Jetted-AGNs}\label{sec:physics}
BL Lacs show an entirely jet-dominated continuum and thus, they are widely
accepted to lack the standard accretion disk and the putative IR-torus\footnote{Recently,
\citet{2021icrc.confE..804K} reported an IR-torus for the first-time in a BL Lac
source, but it is located at much longer distance from the central SMBH than the scale
expected in the standard AGN paradigm. Hence, the sighting of a sharp break in the
NIR-optical spectrum \citep[relative to jet broadband emission;][]{2018MNRAS.479.1672K, 2021ApJ...921...18K,2020MNRAS.498.5424R} as well as a Seyfert-like soft X-ray excess \citep{2020ApJ...890...47P}, and the iron line absorption  feature} \citep{2020MNRAS.498L..35K} indicate OJ 287 as a peculiar BL Lac object. Such diverse
spectral behaviors demand strong changes in physical conditions and an extremely
efficient emission process with radiative output comparable to  that of the  jet/accretion. 


\subsection{Accretion Physics}
Figure \ref{fig:fig2} presents a glimpse of the diverse NIR-optical spectra exhibited
by OJ  287. Panel (a) shows the NIR-optical spectra considered typical of OJ 287,
(b) shows the sharp break between Nthe IR-optical spectra -- first reported by \citet{2018MNRAS.473.1145K}, (c) shows an yet unseen enigmatic variations in the NIR 
K-band data from the SMARTS facility, preceding the appearance of the spectral
break\footnote{The variability is seen only in the SMARTS facility K-band data and is
not in sync with variability seen in other NIR-optical bands and neither with the NIR data
from INAOE, Mexico \citep{2022ApJS..260...39G}, and thus, very very unusual. Further,
it seems to persist for almost one season of observation from the SMARTS facility (MJD:
$\rm\sim$ 55500 -- 55715)}, and (d) shows the broadband SEDs of a flaring and a
quiescent (end) phase from the 2015--2016 MW activity. The  NIR-optical spectral
break  is quite sharp (\(\alpha_{IJ}^{max} \sim -2.5 ; F_\nu ~\sim \nu^{-\alpha}\))
and is inconsistent with  a (smooth) power-law shape that is the typical spectrum
of the source (as well as BL Lacs). Such a sharp break is indicative of a thermal feature -- either thermal emission or a transition from optically thick to a thin  
emission regime. In addition, there is no indication of any  shifting in the location
of this spectral break  throughout the activity period. 

 As stated above, the close coincidence of QPOOs and the spectral changes is
indicative of a link between the two \citep[e.g.][]{2020Galax...8...15K} -- also
supported by limited records available in earlier studies \citep[e.g.][and references
therein]{2001PASJ...53...79I}. Such strong spectral changes are inconsistent
with the simple jet-precession scenarios \citep[and references therein]{2018MNRAS.478.3199B,
2020Univ....6..191B} invoked for the QPOOs in which only achromatic flux boosting
is expected without any spectral changes. If this interpretation is indeed the case,
it requires strong dynamical mechanisms with very efficient radiative output -- like
that of accretion or jet to account for these spectral features with dynamical forces
peaking around the expected QPOOs to drive the spectral changes. On the other hand,
the thermal-like NIR-optical spectral break \citep{2018MNRAS.473.1145K,2021ApJ...921...18K},
Seyfert-like soft X-ray excess \citep{2020ApJ...890...47P}, and iron line absorption
feature \citep{2020MNRAS.498L..35K} are broadly consistent with the elements of the
disk-impact BBH interpretation -- the NIR-spectral break with a \(\sim 10^{10}~M_\odot\)
SMBH accretion disk spectrum \citep{2018MNRAS.473.1145K,2020Galax...8...15K,
2021ApJ...921...18K} or even with the thermal bremsstrahlung \citep{2020MNRAS.498.5424R}
proposition while Seyfert-like soft X-ray excess \citep{2020ApJ...890...47P} as
well as iron absorption feature \citep{2020MNRAS.498L..35K}, are expected in case
of a standard accretion-disk and outflows.  However, observations do not support
the disk-impact model claim of bremsstrahlung as the driver of QPOOs and so do the
optical polarization trends \citep{2020Galax...8...15K}.

Another interesting spectral behavior related to the $\sim\rm12$-yr QPOOs seems to
be an extremely soft X-ray spectral state \citep[and references therein]{2001PASJ...53...79I,
2017IAUS..324..168K,2020MNRAS.498L..35K,2018MNRAS.479.1672K} with peculiar timing
trends \citep{2018MNRAS.479.1672K,2020MNRAS.498L..35K,2021ApJ...921...18K} and an
HBL-like broadband SED \citep{2018MNRAS.479.1672K,2021ApJ...921...18K}. From primarily
the timing perspective, \citet{2021MNRAS.504.5575K} claim this to be the impact-induced 
jet activity that is proposed in the disk-impact BBH scenario. It should, however,
be noted that the broadband spectral state is very different (HBL-like) than the
well-known LBL state to which the source belongs and the BBH model makes no comment
on spectral state other than jet activity. On the other hand, recently \citet{2021ApJ...920...12H}
claims this spectral state with associated peculiar timing feature as a tidal
disruption event (TDE), likely associated with the secondary SMBH. 
These observations and the theoretical consideration make OJ 287 a unique laboratory
to explore diverse aspects of not only extreme jet physics (ref \S\ref{subsec:jet}) but 
accretion as well as the jet-accretion physics in jetted AGNs.

\subsubsection{Disk-impact binary BBH Model vs Observations}\label{sub:obsBBH}
Though the disk-impact BBH interpretation is favored for the $\rm\sim12$-yr QPOOs,
both from timing, spectral and even polarization, many of the aspects are still
ambiguous and contrary to the model claims. The foremost being the claim that thermal
bremsstrahlung -- an emission with a characteristically different spectral shape
compared to the jet broadband emission \citep[see also][]{2020Galax...8...15K} powers
the QPOOs. The multi-band NIR-optical flaring spectra of 2005, 2006, and the most
recent 2015 activities \citep{2009PASJ...61.1011S,2018MNRAS.473.1145K}, however, do
not support this interpretation \citep[see also][]{2012MNRAS.427...77V}. Additionally,
the 2015 QPOO was the first flare with a true MW coverage from radio to gamma-rays
and show not only optical outburst but also flaring at X-rays and gamma-rays, 
indicating a non-thermal emission component. Not only this, the MeV-GeV spectra are
also very different from the usual source spectra \citep[ref Fig. \ref{fig:fig2}-d;]
[]{2018MNRAS.473.1145K}.

Another perplexing feature is the large systematic optical PA rotation observed
during the  $\sim12$-yr QPOOs \citep{2000A&AS..146..141P,2018MNRAS.473.1145K} despite
a lower PD.  In a thermal-emission-powered outburst,  such a systematic PA
swing is not expected in general.
Further, neither the reported PD observations of all these QPOOs outbursts are
consistent with being lower or close to zero \citep[e.g.][]{2009PASJ...61.1011S}
as expected for a  thermal-emission-powered outburst. All of these indicate additional
dynamical processes in play. Note that a recent work based on this model by \citet{2021MNRAS.503.4400D}
shows that the model can reproduce the radio PA swings. However, the one associated
with thermal outbursts remains  perplexing given the claim of thermal origin. The
observation of the next predicted outburst in 2022 holds key to many of the features
and further insight into the complexity of observed behaviors. Regarding the claim
of the non-thermal soft-X-ray state  (HBL-like component driven MW activities of
2016--2017 and 2020) as the impact-induced/triggered jet activity  by \citet{2020MNRAS.498L..35K} on the basis of timing perspective (both model and observational comparison with  other
sources), it should be noted that the timing comparison is drawn with respect to
the optical outbursts  that occur much later to the impact (e.g. $\rm\sim$ 2.5 years
for 2015 outburst) -- once the  claimed torn blob turns optically thin 
\citep{2018ApJ...866...11D} and thus, taking  QPOOs timing as a proxy
for perturbations in disk seems unjustified  in the context of comparison with
other sources where similar timing features have been seen.

\begin{figure*}[!ht]
\begin{center}
\includegraphics[scale=0.55, angle=0]{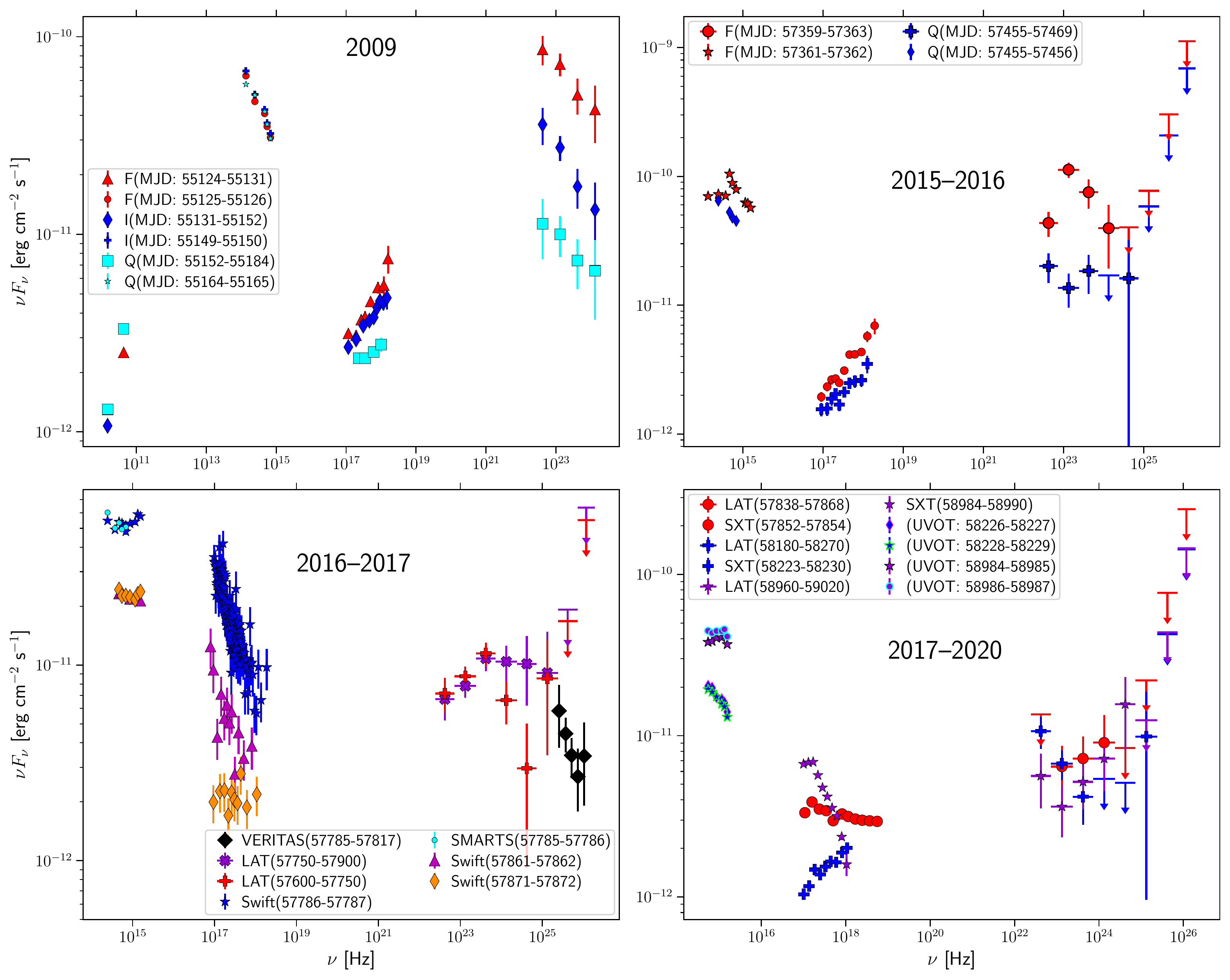}
\end{center}
\caption{A plot summarizing the distinct broadband spectral phases exhibited by OJ 287.
The year-label marked the calendar year of the broadband SEDs. The numbers in the
parenthesis corresponding to the plot label are the MJD duration of the data from 
different facilities while text F, I, and refers to Flaring, Intermediate, and Quiescent
MW flux state of the source. The plot labeled 2009 show the broadband SEDs considered
normal of the source \citep{2013MNRAS.433.2380K} and is the basis spectral classification of the source as LBL from a MW activity in 2009. The plot labeled 2015-2016 show the 
flaring and quiescent SEDs from the 2015--2016 MW activity \citep{2018MNRAS.473.1145K}. 
The 2016--2017 SEDs highlight the strong spectral changes in all bands from NIR-optical
to MeV-GeV and VHE gamma-rays as a result of an HBL-like emission component\citep{2018MNRAS.479.1672K} while the 2017--2020 show the source spectral state from disappearance of 
the HBL-like component in 2017-end to its reappearance in 2020 \citep{2021icrc.confE..644K,2022MNRAS.509.2696S}. The LAT spectra from 2015
onward have been re-analyzed with Fermipy and reproduced here. The VERITAS VHE spectrum
\citep{2017ICRC...35..650B} has been corrected for EBL absorption following \citet{2011MNRAS.410.2556D}. }
\label{fig:fig3}
\end{figure*}

\subsection{Jet Physics}\label{subsec:jet}
The broadband emission and an enormous range of observed behaviors in all the 
observational windows are direct reflections/manifestations of extreme and highly
dynamic physical conditions within the jet. These behaviors are directly related
to the existence of ultra-relativistic non-thermal particles and the associated
plausible emission  channels. The former, in turn, is directly related to the
interplay between magnetic fields and particles, while the latter is related to
the matter constituents of the plasma -- broadly whether primarily leptonic or
hadronic or, if both, in what fractions. 

Peeling the particle acceleration issue further down involves how the evolution
and instabilities within the outflowing plasma led from (probably) magnetic dominated
\citep[e.g.][]{2014Natur.510..126Z} to kinetic-dominated jet as indicated by SED
modeling \citep{2013ApJ...768...54B,2014Natur.515..376G}. For the emission mechanisms,
the fundamental issue remains
is whether the high energy hump is due primarily to primary leptons -- via inverse Compton
scattering, or hadrons -- primarily protons via proton-synchrotron and/or proton-proton,
proton-photon initiated cascades, and if both, the respective contributions \citep[e.g.][]{2017SSRv..207....5R,2019NatAs...3...88G,2018ApJ...865..124M}.

The diverse range of broadband SED exhibited by OJ 287 \citep{2018MNRAS.473.1145K,
2018MNRAS.479.1672K,2021ApJ...921...18K} provide  potential constraints in
exploring different emission scenarios -- within standard blazar emission paradigm
as well as scenarios inspired by QPOOs models.  For the 2015 -- 2016 broadband 
SEDs during quiescent and flaring part (ref Figure \ref{fig:fig1}-d), \citet{2018MNRAS.473.1145K} showed that overall emission can be reproduced in leptonic scenario where
the IC scattering of BLR photons (IC-BLR) is responsible
for the hardening of MeV-GeV spectra and the shift of the high energy peak while the
NIR-optical spectral break can be reproduced by the standard accretion disk spectrum
of a $\rm \sim 10^{10}~M_\odot$ SMBH in the flaring phase. In the quiescent case, the
accretion-disk component is weakened (or disappeared) so much that its no longer
visible while the IC-BLR component is weakened too -- giving rise to a flat MeV-HeV spectrum. This interpretation is consistent with the report of increment of emission
line strength during the previous cycle (2005 -- 2008) of QPOOs \citep[and references
therein]{2010A&A...516A..60N}. \citet{2019MNRAS.489.4347O}, on the other hand,
showed that MeV-GeV spectral hardening can also be reproduced in a hadronic 
scenario by $p\gamma$ channel. However, the NIR-optical break remains unexplained
in the latter case.

In a completely different scenario inspired by disk-impact model of the QPOOs,
\citet{2020MNRAS.498.5424R} showed that the modified broadband SED with the NIR-optical
spectral break and a hardened MeV-GeV spectrum can be self-consistently reproduced
in a non-jetted disk-impact scenario via thermal bremsstrahlung and hadronic pp
cascade, respectively.  This is very different than the standard blazar hadronic
emission scenarios, as it involves no boosting of emission. The model invokes outflows
and is consistent with the report of absorption feature reported at X-ray energies
\citep{2020MNRAS.498L..35K} as well as the Seyfert-like soft X-ray excess \citep{2020ApJ...890...47P}. Energetically too,
the model fares better compared to blazar hadronic emission scenario due to the high density of thermal protons on account of the ejected optically thick blob that makes
the pp interaction effective and thus requires lesser number of ultra-relativistic
protons. In this scenario, it should be noted that the disk-impact BBH model
predicts thermal emission only  for the QPOO (MJD 57361) while the NIR-spectral feature  attributed to thermal bremsstrahlung in this model has been present 
much
before \citep[since MJD $\sim$ 56439;][]{2018MNRAS.473.1145K,2020Galax...8...15K}.

 For the broadband SEDs of the 2016--2017 and 2020 MW activities that show an 
additional emission component similar to HBL, \citet{2018MNRAS.479.1672K} have
shown that the leptonic scenario can reproduce the overall spectrum. In this, the
MeV-GeV is due to the
combined effect of external Comptonization of IR photon and the synchrotron
self-Compton component, where the latter is responsible for hardening of the 
MeV-GeV spectrum \citep[see also ][]{2022MNRAS.509.2696S}. The softening of the 
X-ray and the hardening of the optical-UV spectrum are due to the synchrotron
emission of the HBL component that peaks at UV-soft X-ray region. The explanation
is in line with the phenomenological explanation of the HBL sources and also the
observational fact that the observed X-ray as well as EBL corrected very high
energy (VHE) emission spectrum is similar to the low-state X-ray/VHE spectrum
of the HBLs \citep[and references therein]{2022MNRAS.509.2696S}.

Current studies and constraints suggest mainly a leptonic origin for GeV emission
with a subdominant hadronic emission component \citep[e.g.][]{2018ApJ...865..124M,
2019NatAs...3...88G,2019MNRAS.489.4347O}. In leptonic emission scenario, a highly contentious issue has been the location of the emission region, which is directly
related to
the soft photon field required for an effective inverse Compton scattering. From
studies of the kinematics of superluminal and quasi-stationary features in radio along
with correlation studies with gamma-ray, a parsec scale origin of the emission has
been argued \citep{2011ApJ...726L..13A,2017A&A...597A..80H}. A similar inference
is inferred from broadband SED modeling that requires an IR photon field \citep{2013MNRAS.433.2380K}.  More recent radio studies also indicate a systematic trend in
quasi-stationary knots location \citep{2018MNRAS.478.3199B} -- the claimed location of
emission region or the blazar zone, over a year timescale. So if the knots are the
location of high-energy emission, one expects an energy-independent flux variations
and QPOs in all energy
bands. Timing studies of the Fermi-LAT light curve (0.1 -- 300 GeV) indicate a QPO \citep[but see \citet{2018ApJ...863..175G} and \citep{2020ApJ...896..134P}\footnote{used
$>$ 1 GeV light curve}]{2020MNRAS.499..653K} but low cadence observations in optical
and X-ray bands do not allow such exploration without possible biases. These
considerations combined with spectral properties allow one to locate the emission
region  that has direct implications on issues pertaining scales of energy dissipation and transformation, particle acceleration, etc.

 On the spectrum and emitting particle spectrum front, blazars' broadband emission requires ultra-relativistic particle energies.  Generally, it is believed to be a competitive interplay between the radiative cooling
-- the dominant loss mechanism, and the acceleration process. Typical estimates of
radiative cooling time scales suggest cooling timescales of a fraction of a  minute
\citep[e.g.][]{2014MNRAS.442..131K}  indicating extremely efficient accelerating
mechanisms and extreme physical conditions. Though time-dependent studies have
broadened our general understanding of likely physical conditions \citep{2011ApJ...729...26M,
2014ApJ...789...66Z,2014ApJ...780...87M,2019NatAs...3...88G} and even attempts
have been made to identify physical signatures of underlying acceleration processes
\citep[e.g.][and references therein]{2018ApJ...862L..25Z}, the issue remains the
least understood. The diversity of the spectral changes reported in OJ 287 \citep{2017IAUS..324..168K,2017ICRC...35..650B,2018MNRAS.473.1145K,2018MNRAS.479.1672K,2020MNRAS.498L..35K,2021ApJ...921...18K,2020ApJ...890...47P,2022MNRAS.509.2696S,2021icrc.confE..644K} and the densely monitored MW data combined with polarization properties
\citep{2016ApJ...819L..37V,2017MNRAS.465.4423G,2019AJ....157...95G,2020MNRAS.498L..35K} provide
an excellent source to explore aspects related to particle acceleration processes, acceleration timescales, physical conditions, etc. 

\begin{figure}[!ht]
\begin{center}
\includegraphics[scale=0.55, angle=0]{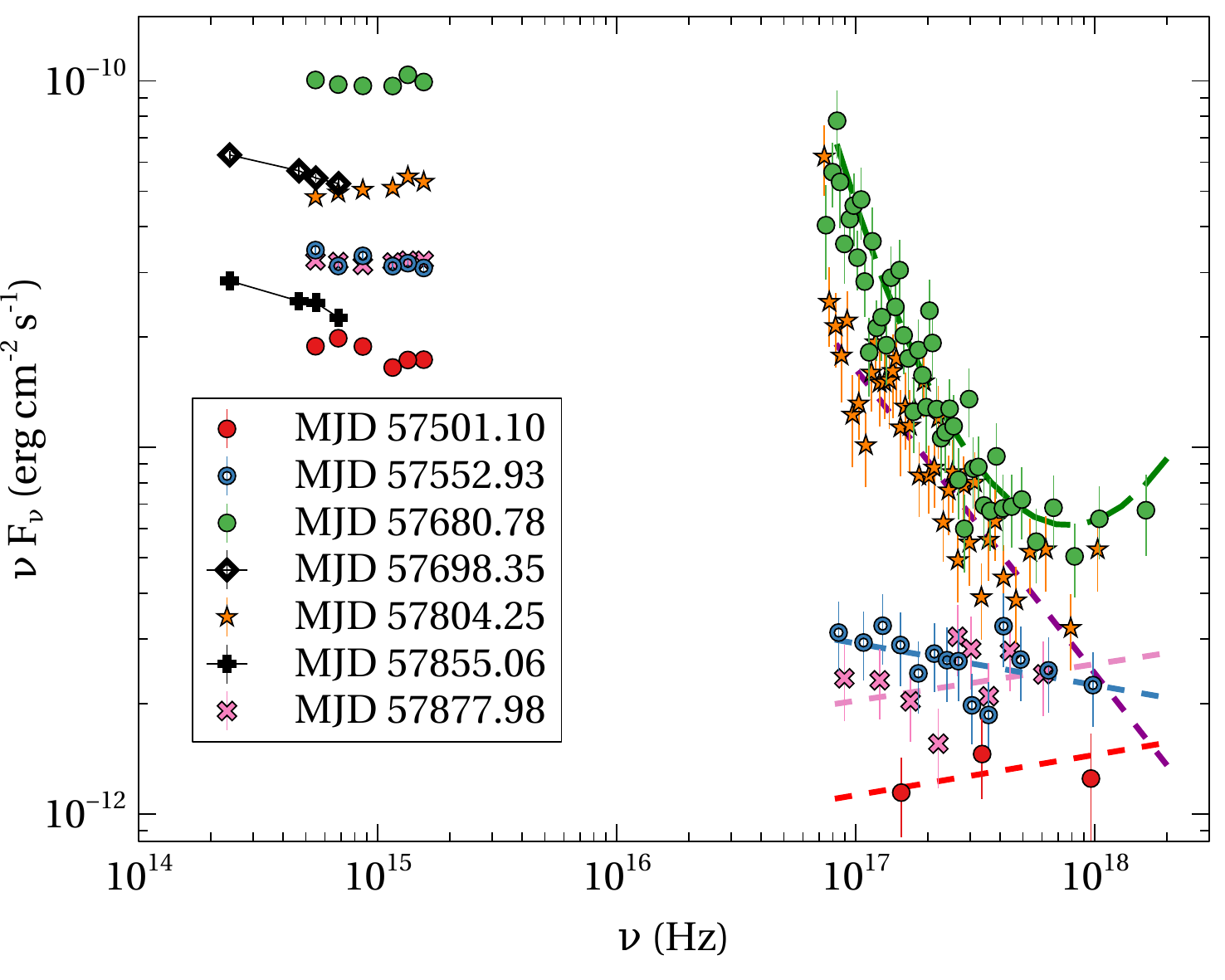}
\end{center}
\caption{A peek into a sequence of diverse optical to X-ray spectral phases exhibited
by OJ 287, highlighting the very dynamic and evolutionary nature of the source spectra.
The optical-X-ray part is the direct tracer of evolution of the high-energy end of the particle spectrum.}
\label{fig:fig4}
\end{figure}

 The other issue related to the emitting particle spectrum is the lowest
and highest achievable particle energies i.e. the two extreme ends of the particle
spectrum.  In this direction, the diverse optical-UV to X-ray spectra of OJ 287
offer excellent inputs\footnote{gamma-ray too, but the weakness of the source 
do not allow short-time evolution history like those of optical to X-rays.} (e.g. Figs. \ref{fig:fig1}, \ref{fig:fig3}, \ref{fig:fig4}),  especially the strong spectral softening/cutoff revealed by the long-exposure observation by {\it AstroSat} during
a low X-ray flux state as reported in \citet{2022MNRAS.509.2696S} (see also \citep{2021icrc.confE..644K}). Under the
blazar paradigm,  it  provides a lower bound on the highest particle energies. 
 The observation of steepening when combined with the general optical-UV and 
X-ray evolution of the source also establishes that most of the X-ray spectral change in
the LBL state of the source is due to evolution of the synchrotron component
\citep{2022MNRAS.509.2696S} which is reflective of the strong evolution in the 
high-energy-end of the underlying particle spectrum.
A similar spectral sharpening/cutoff can be inferred during the low X-ray flux phase of
the HBL-driven MW activity from the spectral shape of the optical-UV and the X-ray spectra
\citep{2018MNRAS.479.1672K,2021ApJ...921...18K,2022MNRAS.509.2696S}. Interestingly, the onset of
spectral steepening is almost at similar energies, indicating a similar acceleration
process or even the same sites for both cases. What is further intriguing is that the
cutoff is seen in the low flux state
of the source. In general, if the radiative loss dominates over accelerating (and often
used in the literature as a proxy for highest particle energies, e.g. \citet{1996ApJ...457..253D}), the cutoff
should be expected during the bright phases. This suggests that the spectral
shape and energies are shaped primarily by the local conditions within the jet
rather than  by cooling alone. Not only this, the observation of HBL-like
emission component in an LBL has implications for the blazar spectral sequence.

On the issue of jet-disk connection and the accretion-jet paradigm, the report
of an additional HBL-like new broadband emission component \citep{2018MNRAS.479.1672K,
2021ApJ...921...18K,2021A&A...654A..38P} with peculiar timing properties \citep{2018MNRAS.479.1672K,2020MNRAS.498L..35K,2021ApJ...921...18K} and claim of this as a TDE 
and/or disk-impact triggered jet activities offer a potential way to explore jet-disk
connection issues, propagation and evolution, and much more.

In short, the multitude of observed features reported in OJ 287 across the directly
accessible observational windows makes it the most promising blazar and jetted-AGNs
to investigate not only jet physics, but accretion as well as accretion-regulated jet
activities. The dense MW monitoring of the most spectrally dramatic activity between
2015 to 2020 when combined with time-dependent modeling and investigations, holds the
potential to deepen our existing understanding. The source is now an EHT (Event 
Horizon Telescope) target,
and the studies and MW observations are expected to significantly broaden our 
existing understanding.

\section{Summary and Conclusions}\label{sec:sumCon}
We presented a brief overview of peculiar features exhibited by OJ 287 since 2013
and implications these context of the standard blazar emission paradigm as well as the scenarios motivated by the proposed model of $\rm \sim 12$-yr QPOO. A summary 
is as follows:

\begin{itemize}
 \item The optical-NIR spectral break is too sharp when seen in the context of the well-known 
 jet spectrum. Its timing and spectral shape broadly favor the disk-impact model over the 
 simple jet precession scenarios. However, many of its predictions/claims are at odds
 with observations e.g. the bremsstrahlung origin of flare and a simultaneous flaring at
 X-ray and gamma-rays, large systematic swing in
 optical PA for a thermal-emission-powered flare (the 2015 and outbursts), etc. Jet 
 precession may be admissible but requires a dynamical model that may allow strong
 spectral changes.
 
 \item Jet precession inferred from radio knots' position and the claim of these knots
 as the origin of high-energy emission provide an additional way to constrain the
 location of emission region through correlated timing studies across the EM bands.
 
 \item The observation of sharp steepening of the high-energy-end of the optical-UV
 spectrum combined with the associated X-ray spectrum during a low-flux state of the source questions the often used assumption constraining the highest achievable energy
 of the particle spectrum by equating to radiative losses. The finding rather indicates local jet conditions as the primary driver. This is also supported by similar steepening
 inferred for the low state of the HBL-like activity. The steepening also settles
 that most of the X-ray spectral evolution in the LBL state of the source
 is due to the synchrotron component i.e. a direct reflection of the evolution of the high-energy end of the underlying particle spectrum.
 
 \item Broadband SED modeling of different spectral states and during different
 activity phases supports the leptonic origin of the MeV-GeV emission with external
 Comptonization as the main driver. The explanation of SEDs of the 2015--2016 MW
 activity in a non-jetted hadronic scenario, supported by many observational clues,
 can be further tested with future observations.
 
 \item The consistency of many elements of the disk-impact model and the claim of an
 HBL-like emission component as TDE offer a potential candidate to explore accretion
 physics and jet-disk connection in addition to jet physics.
\end{itemize}

The MW observations of the expected 2022 QPOO hold additional clues and inputs on these
issues. The diversity of observed peculiar behaviors/trends, their expected time of 
occurrence, and the implications of these on almost all issues pertaining to 
jet-accretion paradigm makes OJ 287 an idea candidate for coordinated MW observations
for further insight on the issues of complex accretion dynamics and jet physics.

\section*{Acknowledgements}
PK acknowledges ARIES A-PDF funding (grant: AO/A-PDF/770). The work has made
use of software, and/or web tools obtained from NASA’s High Energy Astrophysics
Science Archive Research Center (HEASARC), a service of the Goddard Space Flight
Center and the Smithsonian Astrophysical Observatory. This paper has made use of
up-to-date SMARTS optical/near-infrared light curves that are available at
\url{http://www.astro.yale.edu/smarts/glast/home.php}.
\vspace{-1em}




\begin{theunbibliography}{}
\vspace{-1.5em}

\bibitem[Abdo et al.(2010)]{2010ApJ...716...30A} Abdo, A.~A., Ackermann, M., Agudo, I., et al.\ 2010, \apj, 716, 30. doi:10.1088/0004-637X/716/1/30

\bibitem[Agudo et al.(2011)]{2011ApJ...726L..13A} Agudo, I., Jorstad, S.~G., Marscher, A.~P., et al.\ 2011, \apjl, 726, L13. doi:10.1088/2041-8205/726/1/L13

\bibitem[Ahnen et al.(2018)]{2018A&A...620A.181A} Ahnen, M.~L., Ansoldi, S., Antonelli, L.~A., et al.\ 2018, \aap, 620, A181. doi:10.1051/0004-6361/201833704

\bibitem[Anglada et al.(2018)]{2018A&ARv..26....3A} Anglada, G., Rodr{\'\i}guez, L.~F., \& Carrasco-Gonz{\'a}lez, C.\ 2018, \aapr, 26, 3. doi:10.1007/s00159-018-0107-z

\bibitem[Bhattacharyya et al.(2020)]{2020ApJ...897...25B} Bhattacharyya, S., Ghosh, R., Chatterjee, R., et al.\ 2020, \apj, 897, 25. doi:10.3847/1538-4357/ab91a8

\bibitem[Blandford et al.(2019)]{2019ARA&A..57..467B} Blandford, R., Meier, D., \& Readhead, A.\ 2019, \araa, 57, 467. doi:10.1146/annurev-astro-081817-051948

\bibitem[Blinov \& Pavlidou(2019)]{2019Galax...7...46B} Blinov, D. \& Pavlidou, V.\ 2019, Galaxies, 7, 46. doi:10.3390/galaxies7020046

\bibitem[Blinov et al.(2018)]{2018MNRAS.474.1296B} Blinov, D., Pavlidou, V., Papadakis, I., et al.\ 2018, \mnras, 474, 1296. doi:10.1093/mnras/stx2786

\bibitem[B{\"o}ttcher et al.(2013)]{2013ApJ...768...54B} B{\"o}ttcher, M., Reimer, A., Sweeney, K., et al.\ 2013, \apj, 768, 54. doi:10.1088/0004-637X/768/1/54

\bibitem[Boula et al.(2019)]{2019MNRAS.482L..80B} Boula, S., Kazanas, D., \& Mastichiadis, A.\ 2019, \mnras, 482, L80. doi:10.1093/mnrasl/sly189

\bibitem[Brien \& VERITAS Collaboration(2017)]{2017ICRC...35..650B} Brien, S.~O. \& VERITAS Collaboration\ 2017, 35th International Cosmic Ray Conference (ICRC2017), 301, 650

\bibitem[Britzen et al.(2018)]{2018MNRAS.478.3199B} Britzen, S., Fendt, C., Witzel, G., et al.\ 2018, \mnras, 478, 3199. doi:10.1093/mnras/sty1026

\bibitem[Butuzova \& Pushkarev(2020)]{2020Univ....6..191B} Butuzova, M.~S. \& Pushkarev, A.~B.\ 2020, Universe, 6, 191. doi:10.3390/universe6110191

\bibitem[Carini et al.(1992)]{1992AJ....104...15C} Carini, M.~T., Miller, H.~R., Noble, J.~C., et al.\ 1992, \aj, 104, 15. doi:10.1086/116217

\bibitem[Cohen(2017)]{2017Galax...5...12C} Cohen, M.\ 2017, Galaxies, 5, 12. doi:10.3390/galaxies5010012

\bibitem[de Jager et al.(1996)]{1996ApJ...457..253D} de Jager, O.~C., Harding, A.~K., Michelson, P.~F., et al.\ 1996, \apj, 457, 253. doi:10.1086/176726

\bibitem[Dey et al.(2021)]{2021MNRAS.503.4400D} Dey, L., Valtonen, M.~J., Gopakumar, A., et al.\ 2021, \mnras, 503, 4400. doi:10.1093/mnras/stab730

\bibitem[Dey et al.(2018)]{2018ApJ...866...11D} Dey, L., Valtonen, M.~J., Gopakumar, A., et al.\ 2018, \apj, 866, 11. doi:10.3847/1538-4357/aadd95

\bibitem[Dom{\'\i}nguez et al.(2011)]{2011MNRAS.410.2556D} Dom{\'\i}nguez, A., Primack, J.~R., Rosario, D.~J., et al.\ 2011, \mnras, 410, 2556. doi:10.1111/j.1365-2966.2010.17631.x

\bibitem[Event Horizon Telescope Collaboration et al.(2019)]{2019ApJ...875L...1E} Event Horizon Telescope Collaboration, Akiyama, K., Alberdi, A., et al.\ 2019, \apjl, 875, L1. doi:10.3847/2041-8213/ab0ec7

\bibitem[Event Horizon Telescope Collaboration et al.(2019)]{2019ApJ...875L...5E} Event Horizon Telescope Collaboration, Akiyama, K., Alberdi, A., et al.\ 2019, \apjl, 875, L5. doi:10.3847/2041-8213/ab0f43

\bibitem[Finke(2013)]{2013ApJ...763..134F} Finke, J.~D.\ 2013, \apj, 763, 134. doi:10.1088/0004-637X/763/2/134

\bibitem[Gao et al.(2019)]{2019NatAs...3...88G} Gao, S., Fedynitch, A., Winter, W., et al.\ 2019, Nature Astronomy, 3, 88. doi:10.1038/s41550-018-0610-1

\bibitem[Ghisellini et al.(2017)]{2017MNRAS.469..255G} Ghisellini, G., Righi, C., Costamante, L., et al.\ 2017, \mnras, 469, 255. doi:10.1093/mnras/stx806

\bibitem[Ghisellini et al.(2014)]{2014Natur.515..376G} Ghisellini, G., Tavecchio, F., Maraschi, L., et al.\ 2014, \nat, 515, 376. doi:10.1038/nature13856

\bibitem[Goyal(2021)]{2021ApJ...909...39G} Goyal, A.\ 2021, \apj, 909, 39. doi:10.3847/1538-4357/abd7fb

\bibitem[Goyal et al.(2018)]{2018ApJ...863..175G} Goyal, A., Stawarz, {\L}., Zola, S., et al.\ 2018, \apj, 863, 175. doi:10.3847/1538-4357/aad2de

\bibitem[Granot \& van der Horst(2014)]{2014PASA...31....8G} Granot, J. \& van der Horst, A.~J.\ 2014, \pasa, 31, e008. doi:10.1017/pasa.2013.44

\bibitem[Gupta et al.(2017)]{2017MNRAS.465.4423G} Gupta, A.~C., Agarwal, A., Mishra, A., et al.\ 2017, \mnras, 465, 4423. doi:10.1093/mnras/stw3045

\bibitem[Gupta et al.(2019)]{2019AJ....157...95G} Gupta, A.~C., Gaur, H., Wiita, P.~J., et al.\ 2019, \aj, 157, 95. doi:10.3847/1538-3881/aafe7d

\bibitem[Gupta et al.(2022)]{2022ApJS..260...39G} Gupta, A.~C., Kushwaha, P., Carrasco, L., et al.\ 2022, \apjs, 260, 39. doi:10.3847/1538-4365/ac6c2c

\bibitem[Hayashida et al.(2015)]{2015ApJ...807...79H} Hayashida, M., Nalewajko, K., Madejski, G.~M., et al.\ 2015, \apj, 807, 79. doi:10.1088/0004-637X/807/1/79

\bibitem[Hervet et al.(2016)]{2016A&A...592A..22H} Hervet, O., Boisson, C., \& Sol, H.\ 2016, \aap, 592, A22. doi:10.1051/0004-6361/201628117

\bibitem[Hodgson et al.(2017)]{2017A&A...597A..80H} Hodgson, J.~A., Krichbaum, T.~P., Marscher, A.~P., et al.\ 2017, \aap, 597, A80. doi:10.1051/0004-6361/201526727

\bibitem[Huang et al.(2021)]{2021ApJ...920...12H} Huang, S., Hu, S., Yin, H., et al.\ 2021, \apj, 920, 12. doi:10.3847/1538-4357/ac0eff

\bibitem[Hudec et al.(2013)]{2013A&A...559A..20H} Hudec, R., Ba{\v{s}}ta, M., Pihajoki, P., et al.\ 2013, \aap, 559, A20. doi:10.1051/0004-6361/201219323

\bibitem[Isobe et al.(2001)]{2001PASJ...53...79I} Isobe, N., Tashiro, M., Sugiho, M., et al.\ 2001, \pasj, 53, 79. doi:10.1093/pasj/53.1.79

\bibitem[Joshi \& B{\"o}ttcher(2011)]{2011ApJ...727...21J} Joshi, M. \& B{\"o}ttcher, M.\ 2011, \apj, 727, 21. doi:10.1088/0004-637X/727/1/21

\bibitem[Kapanadze et al.(2018)]{2018MNRAS.480..407K} Kapanadze, B., Vercellone, S., Romano, P., et al.\ 2018, \mnras, 480, 407. doi:10.1093/mnras/sty1803

\bibitem[Kiehlmann et al.(2016)]{2016A&A...590A..10K} Kiehlmann, S., Savolainen, T., Jorstad, S.~G., et al.\ 2016, \aap, 590, A10. doi:10.1051/0004-6361/201527725

\bibitem[Murase \& Bartos(2019)]{2019ARNPS..69..477M} Murase, K. \& Bartos, I.\ 2019, Annual Review of Nuclear and Particle Science, 69, 477. doi:10.1146/annurev-nucl-101918-023510

\bibitem[Komossa et al.(2021)]{2021MNRAS.504.5575K} Komossa, S., Grupe, D., Parker, M.~L., et al.\ 2021, \mnras, 504, 5575. doi:10.1093/mnras/stab1223

\bibitem[Komossa et al.(2020)]{2020MNRAS.498L..35K} Komossa, S., Grupe, D., Parker, M.~L., et al.\ 2020, \mnras, 498, L35. doi:10.1093/mnrasl/slaa125

\bibitem[Komossa et al.(2017)]{2017IAUS..324..168K} Komossa, S., Grupe, D., Schartel, N., et al.\ 2017, New Frontiers in Black Hole Astrophysics, 324, 168. doi:10.1017/S1743921317001648

\bibitem[Kushwaha(2020)]{2020Galax...8...15K} Kushwaha, P.\ 2020, Galaxies, 8, 15. doi:10.3390/galaxies8010015

\bibitem[Kushwaha \& Pal(2020)]{2020Galax...8...66K} Kushwaha, P. \& Pal, M.\ 2020, Galaxies, 8, 66. doi:10.3390/galaxies8030066

\bibitem[Kushwaha et al.(2016)]{2016ApJ...822L..13K} Kushwaha, P., Chandra, S., Misra, R., et al.\ 2016, \apjl, 822, L13. doi:10.3847/2041-8205/822/1/L13

\bibitem[Kushwaha et al.(2018a)]{2018MNRAS.473.1145K} Kushwaha, P., Gupta, A.~C., Wiita, P.~J., et al.\ 2018a, \mnras, 473, 1145. doi:10.1093/mnras/stx2394

\bibitem[Kushwaha et al.(2018b)]{2018MNRAS.479.1672K} Kushwaha, P., Gupta, A.~C., Wiita, P.~J., et al.\ 2018b, \mnras, 479, 1672. doi:10.1093/mnras/sty1499

\bibitem[Kushwaha et al.(2021)]{2021ApJ...921...18K} Kushwaha, P., Pal, M., Kalita, N., et al.\ 2021, \apj, 921, 18. doi:10.3847/1538-4357/ac19b8

\bibitem[Kushwaha et al.(2014)]{2014MNRAS.442..131K} Kushwaha, P., Sahayanathan, S., Lekshmi, R., et al.\ 2014, \mnras, 442, 131. doi:10.1093/mnras/stu836

\bibitem[Kushwaha et al.(2013)]{2013MNRAS.433.2380K} Kushwaha, P., Sahayanathan, S., \& Singh, K.~P.\ 2013, \mnras, 433, 2380. doi:10.1093/mnras/stt904

\bibitem[Kushwaha et al.(2020)]{2020MNRAS.499..653K} Kushwaha, P., Sarkar, A., Gupta, A.~C., et al.\ 2020, \mnras, 499, 653. doi:10.1093/mnras/staa2899

\bibitem[Kushwaha et al.(2017)]{2017ApJ...849..138K} Kushwaha, P., Sinha, A., Misra, R., et al.\ 2017, \apj, 849, 138. doi:10.3847/1538-4357/aa8ef5

\bibitem[Kushwaha et al.(2021)]{2021icrc.confE..644K} Kushwaha, P., Singh, K.~P., Sinha, A., et al.\ 2021, 37th International Cosmic Ray Conference {\textemdash} PoS(ICRC2021). 12-23 July 2021, 395, 644. doi:10.22323/1.395.0644

\bibitem[Laine et al.(2020)]{2020ApJ...894L...1L} Laine, S., Dey, L., Valtonen, M., et al.\ 2020, \apjl, 894, L1. doi:10.3847/2041-8213/ab79a4

\bibitem[Lehto \& Valtonen(1996)]{1996ApJ...460..207L} Lehto, H.~J. \& Valtonen, M.~J.\ 1996, \apj, 460, 207. doi:10.1086/176962

\bibitem[Liodakis \& Petropoulou(2020)]{2020ApJ...893L..20L} Liodakis, I. \& Petropoulou, M.\ 2020, \apjl, 893, L20. doi:10.3847/2041-8213/ab830a

\bibitem[Lister et al.(2013)]{2013AJ....146..120L} Lister, M.~L., Aller, M.~F., Aller, H.~D., et al.\ 2013, \aj, 146, 120. doi:10.1088/0004-6256/146/5/120

\bibitem[Liu et al.(2021)]{2021Univ....7...15L} Liu, X., Wang, X., Chang, N., et al.\ 2021, Universe, 7, 15. doi:10.3390/universe7010015

\bibitem[Marscher(2014)]{2014ApJ...780...87M} Marscher, A.~P.\ 2014, \apj, 780, 87. doi:10.1088/0004-637X/780/1/87

\bibitem[Marscher et al.(2010)]{2010ApJ...710L.126M} Marscher, A.~P., Jorstad, S.~G., Larionov, V.~M., et al.\ 2010, \apjl, 710, L126. doi:10.1088/2041-8205/710/2/L126

\bibitem[Marscher \& Jorstad(2011)]{2011ApJ...729...26M} Marscher, A.~P. \& Jorstad, S.~G.\ 2011, \apj, 729, 26. doi:10.1088/0004-637X/729/1/26

\bibitem[Murase et al.(2018)]{2018ApJ...865..124M} Murase, K., Oikonomou, F., \& Petropoulou, M.\ 2018, \apj, 865, 124. doi:10.3847/1538-4357/aada00

\bibitem[Nilsson et al.(2010)]{2010A&A...516A..60N} Nilsson, K., Takalo, L.~O., Lehto, H.~J., et al.\ 2010, \aap, 516, A60. doi:10.1051/0004-6361/201014198

\bibitem[Oikonomou et al.(2019)]{2019MNRAS.489.4347O} Oikonomou, F., Murase, K., Padovani, P., et al.\ 2019, \mnras, 489, 4347. doi:10.1093/mnras/stz2246

\bibitem[Pal et al.(2020)]{2020ApJ...890...47P} Pal, M., Kushwaha, P., Dewangan, G.~C., et al.\ 2020, \apj, 890, 47. doi:10.3847/1538-4357/ab65ee

\bibitem[Pe{\~n}il et al.(2020)]{2020ApJ...896..134P} Pe{\~n}il, P., Dom{\'\i}nguez, A., Buson, S., et al.\ 2020, \apj, 896, 134. doi:10.3847/1538-4357/ab910d

\bibitem[Petropoulou \& Mastichiadis(2015)]{2015MNRAS.447...36P} Petropoulou, M. \& Mastichiadis, A.\ 2015, \mnras, 447, 36. doi:10.1093/mnras/stu2364

\bibitem[Pian et al.(1998)]{1998ApJ...492L..17P} Pian, E., Vacanti, G., Tagliaferri, G., et al.\ 1998, \apjl, 492, L17. doi:10.1086/311083

\bibitem[Potter \& Cotter(2012)]{2012MNRAS.423..756P} Potter, W.~J. \& Cotter, G.\ 2012, \mnras, 423, 756. doi:10.1111/j.1365-2966.2012.20918.x

\bibitem[Prince et al.(2021a)]{2021A&A...654A..38P} Prince, R., Agarwal, A., Gupta, N., et al.\ 2021a, \aap, 654, A38. doi:10.1051/0004-6361/202140708

\bibitem[Prince et al.(2021b)]{2021MNRAS.508..315P} Prince, R., Raman, G., Khatoon, R., et al.\ 2021b, \mnras, 508, 315. doi:10.1093/mnras/stab2545


\bibitem[Raiteri et al.(2017)]{2017Natur.552..374R} Raiteri, C.~M., Villata, M., Acosta-Pulido, J.~A., et al.\ 2017, \nat, 552, 374. doi:10.1038/nature24623

\bibitem[Raiteri et al.(2007)]{2007A&A...473..819R} Raiteri, C.~M., Villata, M., Larionov, V.~M., et al.\ 2007, \aap, 473, 819. doi:10.1051/0004-6361:20078289

\bibitem[Rodr{\'\i}guez-Ram{\'\i}rez et al.(2021)]{2021icrc.confE..804R} Rodr{\'\i}guez-Ram{\'\i}rez, J.~C., de Gouveia Dal Pino, E.~M., Rafael, A.~B., et al.\ 2021, 37th International Cosmic Ray Conference {\textemdash} PoS(ICRC2021). 12-23 July 2021, 395, 1016. doi:10.22323/1.395.1016

\bibitem[Rodr{\'\i}guez-Ram{\'\i}rez et al.(2020)]{2020MNRAS.498.5424R} Rodr{\'\i}guez-Ram{\'\i}rez, J.~C., Kushwaha, P., de Gouveia Dal Pino, E.~M., et al.\ 2020, \mnras, 498, 5424. doi:10.1093/mnras/staa2664

\bibitem[Romero et al.(2017)]{2017SSRv..207....5R} Romero, G.~E., Boettcher, M., Markoff, S., et al.\ 2017, \ssr, 207, 5. doi:10.1007/s11214-016-0328-2

\bibitem[Roychowdhury et al.(2021)]{2021icrc.confE..804K} Roychowdhury, A., Eileen, E.~T.,  Georganopoulos, M., et al.\ 2021, 37th International Cosmic Ray Conference {\textemdash} PoS(ICRC2021). 12-23 July 2021, 395, 804. doi:10.22323/1.395.0804 

\bibitem[Seta et al.(2009)]{2009PASJ...61.1011S} Seta, H., Isobe, N., Tashiro, M.~S., et al.\ 2009, \pasj, 61, 1011. doi:10.1093/pasj/61.5.1011

\bibitem[Shukla et al.(2018)]{2018ApJ...854L..26S} Shukla, A., Mannheim, K., Patel, S.~R., et al.\ 2018, \apjl, 854, L26. doi:10.3847/2041-8213/aaacca

\bibitem[Sillanpaa et al.(1988)]{1988ApJ...325..628S} Sillanpaa, A., Haarala, S., Valtonen, M.~J., et al.\ 1988, \apj, 325, 628. doi:10.1086/166033

\bibitem[Singh et al.(2022)]{2022MNRAS.509.2696S} Singh, K.~P., Kushwaha, P., Sinha, A., et al.\ 2022, \mnras, 509, 2696. doi:10.1093/mnras/stab3161


\bibitem[Sinha et al.(2018)]{2018MNRAS.480L.116S} Sinha, A., Khatoon, R., Misra, R., et al.\ 2018, \mnras, 480, L116. doi:10.1093/mnrasl/sly136

\bibitem[Sitko \& Junkkarinen(1985)]{1985PASP...97.1158S} Sitko, M.~L. \& Junkkarinen, V.~T.\ 1985, \pasp, 97, 1158. doi:10.1086/131679

\bibitem[Sobolewska et al.(2014)]{2014ApJ...786..143S} Sobolewska, M.~A., Siemiginowska, A., Kelly, B.~C., et al.\ 2014, \apj, 786, 143. doi:10.1088/0004-637X/786/2/143

\bibitem[Stickel et al.(1989)]{1989A&AS...80..103S} Stickel, M., Fried, J.~W., \& Kuehr, H.\ 1989, \aaps, 80, 103

\bibitem[Stickel et al.(1991)]{1991ApJ...374..431S} Stickel, M., Padovani, P., Urry, C.~M., et al.\ 1991, \apj, 374, 431. doi:10.1086/170133

\bibitem[Pursimo et al.(2000)]{2000A&AS..146..141P} Pursimo, T., Takalo, L.~O., Sillanp{\"a}{\"a}, A., et al.\ 2000, \aaps, 146, 141. doi:10.1051/aas:2000264

\bibitem[Tavecchio et al.(2011)]{2011A&A...534A..86T} Tavecchio, F., Becerra-Gonzalez, J., Ghisellini, G., et al.\ 2011, \aap, 534, A86. doi:10.1051/0004-6361/201117204

\bibitem[Tavecchio et al.(2020)]{2020MNRAS.497.1294T} Tavecchio, F., Bonnoli, G., \& Galanti, G.\ 2020, \mnras, 497, 1294. doi:10.1093/mnras/staa2055

\bibitem[Tetarenko et al.(2016)]{2016ApJS..222...15T} Tetarenko, B.~E., Sivakoff, G.~R., Heinke, C.~O., et al.\ 2016, \apjs, 222, 15. doi:10.3847/0067-0049/222/2/15

\bibitem[Uchiyama et al.(2006)]{2006ApJ...648..910U} Uchiyama, Y., Urry, C.~M., Cheung, C.~C., et al.\ 2006, \apj, 648, 910. doi:10.1086/505964

\bibitem[Valtonen et al.(2012)]{2012MNRAS.427...77V} Valtonen, M.~J., Ciprini, S., \& Lehto, H.~J.\ 2012, \mnras, 427, 77. doi:10.1111/j.1365-2966.2012.21861.x

\bibitem[Valtonen et al.(2016)]{2016ApJ...819L..37V} Valtonen, M.~J., Zola, S., Ciprini, S., et al.\ 2016, \apjl, 819, L37. doi:10.3847/2041-8205/819/2/L37

\bibitem[Visvanathan \& Elliot(1973)]{1973ApJ...179..721V} Visvanathan, N. \& Elliot, J.~L.\ 1973, \apj, 179, 721. doi:10.1086/151911

\bibitem[Zamaninasab et al.(2014)]{2014Natur.510..126Z} Zamaninasab, M., Clausen-Brown, E., Savolainen, T., et al.\ 2014, \nat, 510, 126. doi:10.1038/nature13399

\bibitem[Zdziarski et al.(2015)]{2015MNRAS.451..927Z} Zdziarski, A.~A., Sikora, M., Pjanka, P., et al.\ 2015, \mnras, 451, 927. doi:10.1093/mnras/stv986

\bibitem[Zhang et al.(2014)]{2014ApJ...789...66Z} Zhang, H., Chen, X., \& B{\"o}ttcher, M.\ 2014, \apj, 789, 66. doi:10.1088/0004-637X/789/1/66

\bibitem[Zhang et al.(2018)]{2018ApJ...862L..25Z} Zhang, H., Li, X., Guo, F., et al.\ 2018, \apjl, 862, L25. doi:10.3847/2041-8213/aad54f


\end{theunbibliography}
\end{document}